\documentclass[pra,twocolumn,floatfix,nofootinbib]{revtex4-1}
\usepackage{amsmath,amssymb,amsthm}
\usepackage{graphicx} 
\usepackage{epstopdf}
\usepackage{bbm}
\usepackage[colorlinks,allcolors=blue]{hyperref}
\usepackage[up]{subfigure}

\usepackage{commath}

\makeatother

\begin{document}

\title{Quantum secret sharing for a multipartite system under energy dissipation}

\author{Siddh$\bar{a}$nt Singh$^{1,a}$}
\email{siddhant.singh@iitkgp.ac.in}
\author{Shivang Srivastava$^{1,b}$}
\email{sri.shivang@gmail.com}
\affiliation{$^1$\textit{\textbf{{These authors have contributed equally to this work (shared first authorship).}}}}
\affiliation{$^a$Department of Physics, Indian Institute of Technology Kharagpur, Kharagpur 721302, India}
\affiliation{$^b$Department of Physics, Birla Institute of Technology Mesra, Ranchi 835215, India}

\author{Prasanta K Panigrahi}
\email{pprasanta@iiserkol.ac.in}
\affiliation{Department of Physical Sciences,\\ Indian Institute of Science Education and Research Kolkata, West Bengal 741246, India }

\begin{abstract}
We propose a protocol for multipartite secret sharing of quantum information through an \textit{amplitude damping} quantum channel. This network is, for example, of two organizations communicating with their own employees connected via classical channels locally. We  consider a GHZ state distributed among four members in an asymmetric fashion where the members of a sub-party collaborate to decode the received information at their end. The target is to send two bits of information in \textit{one execution} of the protocol. Firstly, we consider an ideal channel and observe that our protocol enables decoding of a secret 2-bit information with unit probability. This is accomplished by one of the senders by the use of a globally operated \textit{quantum teleportation operator}. Secondly, we implement the same protocol in a realistic scenario under energy dissipation by the use of a parameterized \textit{amplitude damping channel} with variable noise. This noise is associated with energy dissipation and hence, loss of probability to distinguish and decode the information at the receiving end. Finally, we make this task possible through an optimization algorithm. Various channel 
\textit{quality measures} are also quantitatively ascertained.

\end{abstract}

\pacs{Valid PACS appear here}
\maketitle

\section{Introduction}
\label{sec:introduction}
Secret sharing is a scheme in cryptography in which a secret information is split among a group of people, who decode the secret by sharing their classical results with each other and combining them in some possible way. It was first discovered by Adi Shamir \cite{ref1} and George Blakley independently in 1979 and later in 1998, this idea was applied in quantum domain by Mark Hillery, Vladimir Buzek and Andre Berthiaume \cite{ref2}. It is achieved by distributing a maximally entangled state such as the GHZ,  among the parties. However, secret sharing without the use of entanglement has also been shown \cite{ref3}.

Previously, the protocol for secret sharing through noisy quantum channels was designed for three qubit GHZ state distributed among three members, specifically for the \textit{phase-damping} channel \cite{ref4}. Several attempts have been made to realize the scheme in a more practical manner which considered various entanglement resources such as maximally entangled, partially entangled \cite{ref5} and entanglement-bereft \cite{ref3} quantum states. The fundamental issue in quantum communication and quantum cryptography is the use of quantum channels. Qubits require a quantum channel to transform without being a subject to quantum decoherence \cite{ref6} and quantum collapse which destroys their quantum nature. Quantum channels are quite beneficial as they can carry quantum information as well as classical information \cite{ref7}. Mathematically, these channels are convex-linear, positive-definite unitary operators \cite{ref8}. We use the \textit{operator-sum-representation} or \textit{Kraus-representation }of quantum channels in this paper \cite{ref9}. Realistic quantum channels are subject to noise and open system, non-unitary dynamics as they lose coherence and entanglement which are the advantage of using quantum computation over the classical information processing. To still make the formalism unitary and hermitian, we first evolve the system and the environment together as a closed system and then trace out the environment after the action of the channel to look at the dynamics of the system alone under the action of the noisy channels \cite{ref9}.

We basically have three types of quantum channels (depolarizing, dephasing and amplitude damping) \cite{ref10} in presence of environment while dealing with noisy operations on qubits. Other noisy channels can be realized as composition of these three channels. This paper concerns with energy dissipating \textit{amplitude-damping} \cite{ref9,ref10} noisy quantum channel as it is the most practical one. The foremost thing which can be imagined for a noisy channel is the energy loss of the information and applying it to multipartite secret sharing becomes more realistic. Since there is a loss of information through a noisy channel, we lose the probability to distinguish between the states undergoing through such a channel. We thus need to know the possible states, then use a POVM (Positive Operator Valued Measurement) to optimize them and maximize the probability to distinguish as done in \cite{ref4}.

In our protocol, two parties are involved each having two members. These parties collaborate with each other and within themselves to make the protocol successful. The protocol is designed in such a way that a party receives two qubits each from an external source and one of those parties gets qubits symmetrically (the sender) while for the other, one of the members gets two (the receiver). The first party members perform local operations on their qubits independent of one another (this sequence of operations is the secret to be sent). After doing this, they send both of their qubits to one of the members of the second party, who collaborate together to decode this secret information. This is assisted by sharing one classical bit from first party to second one to reveal one of their operations and make unique identification, as we show in next section. When we implement the same in a noisy environment, the output states are mixed and non-orthogonal, thus cannot be perfectly distinguished. We then need an optimization algorithm for the decryption at the cost of losing some probability to distinguish the mixed states. 

The paper is organized as follows. In Section-\ref{sec:ideal}, we propose the protocol for an ideal quantum channel without noise. In Section-\ref{sec:noisy protocol}, we implement the protocol for a noisy channel (with energy dissipating nature - amplitude damping) and Section-\ref{sec:quality} is devoted to \textit{quality measures} of the noisy channel. 

\section{Secret sharing without noise}
\label{sec:ideal}
The protocol describes the scenario of communication between two organizations (parties) each having two employees. Party-1 has members A \& B and party-2 has C \& D. The aim of the protocol is to  first consider a GHZ state,
\begin{equation}
|GHZ\rangle=\frac{|0000\rangle+|1111\rangle}{\sqrt[]{2}}
\end{equation}
prepared by an outside party S which acts as a state preparation party and, as an entanglement resource generator for the network to communicate. The GHZ state is a maximally entangled state in $\mathbb{C}^{16}$. It is distributed asymmetrically between two parties, one qubit each is sent to Alice (A) \& Bob (B) while two qubits are sent to Dennis (D) via a quantum channel. This is already pre-decided mutually by the members. Now, A \& B measure their qubits locally and send it to Charlie (C) via quantum channel. He has to decode the operations carried out by A \& B, and while doing this, he must collaborate with Dennis to determine the operations with maximum probability. Charlie also receives a quantum mechanically encoded classical bit as a quantum state describing the operation performed by Bob. This is made possible by a \textit{quantum teleportation} scheme \cite{ref11} which we will show. This protocol consists of following steps:
\subsubsection*{\textbf{Step-1: The GHZ state shared by A, B \& D}}
\label{step1}
\begin{equation}
|GHZ\rangle_{ABDD}=\frac{|0000\rangle+|1111\rangle}{\sqrt[]{2}}
\end{equation}
\subsubsection*{\textbf{Step-2: Alice performs operations on her qubit}}
\label{step2}
Now, Alice performs the set of unitary operations $\{I,\sigma_x,i\sigma_y,\sigma_z\}$ which give the possibility of four states:
\begin{equation}
\begin{split}
I: |\lambda\rangle_{ABDD}=\frac{|0000\rangle+|1111\rangle}{\sqrt[]{2}}\\
\sigma_x: |\lambda\rangle_{ABDD}=\frac{|1000\rangle+|0111\rangle}{\sqrt[]{2}}\\
\sigma_{iy}: |\lambda\rangle_{ABDD}=\frac{|1000\rangle-|0111\rangle}{\sqrt[]{2}}\\
\sigma_{z}: |\lambda\rangle_{ABDD}=\frac{|0000\rangle-|1111\rangle}{\sqrt[]{2}}\\
\end{split}
\end{equation}
\subsubsection*{\textbf{Step-3: Bob performs operations on his qubit}}
\label{step3}
Now, Bob performs the set of unitary operations $\{I,\sigma_x,i\sigma_y,\sigma_z\}$ on qubit in  his possession, which gives the possibility of 16 states in principle, having pairwise similarity and leaves us with 8 possible states ($(a\otimes b)$ meaning that Alice performs $a$ and Bob performs $b$):
\begin{equation}
\begin{split}
\{(I\otimes I),(\sigma_z\otimes \sigma_z)\}: |\lambda\rangle_{ABDD}=\frac{|0000\rangle+|1111\rangle}{\sqrt[]{2}}\\
\{(I\otimes \sigma_z),(\sigma_z\otimes I)\}: |\lambda\rangle_{ABDD}=\frac{|0000\rangle-|1111\rangle}{\sqrt[]{2}}\\
\{(I\otimes \sigma_x),(\sigma_z\otimes i\sigma_y)\}: |\lambda\rangle_{ABDD}=\frac{|0100\rangle+|1011\rangle}{\sqrt[]{2}}\\
\{(I\otimes i\sigma_y),(\sigma_z\otimes \sigma_x)\}: |\lambda\rangle_{ABDD}=\frac{|0100\rangle-|1011\rangle}{\sqrt[]{2}}\\
\{(\sigma_x\otimes I),(i\sigma_y\otimes \sigma_z)\}: |\lambda\rangle_{ABDD}=\frac{|1000\rangle+|0111\rangle}{\sqrt[]{2}}\\
\{(\sigma_x\otimes \sigma_z),(i\sigma_y\otimes I)\}: |\lambda\rangle_{ABDD}=\frac{|1000\rangle-|0111\rangle}{\sqrt[]{2}}\\
\{(\sigma_x\otimes \sigma_x),(i\sigma_y\otimes i\sigma_y)\}: |\lambda\rangle_{ABDD}=\frac{|1100\rangle+|0011\rangle}{\sqrt[]{2}}\\
\{(\sigma_x\otimes i\sigma_y),(i\sigma_y\otimes \sigma_x)\}: |\lambda\rangle_{ABDD}=\frac{|1100\rangle-|0011\rangle}{\sqrt[]{2}}\\
\end{split}
\end{equation}
Notice the symmetry between the states. Here, two sequences of operations performed by Alice and Bob lead us to the same global state. A unique sequence of operations by Alice and Bob can be known, if and only if one of them declares which operation he/she applied, and we choose Bob to perform this task later on. These states may be re-expressed in terms of the Bell-basis as:
\begin{equation}
\begin{split}
\{(I\otimes I),(\sigma_z\otimes \sigma_z)\}: |\lambda\rangle_{1}=\frac{|\Phi^+\rangle|\Phi^+\rangle+|\Phi^-\rangle|\Phi^-\rangle}{2}\\
\{(I\otimes \sigma_z),(\sigma_z\otimes I)\}: |\lambda\rangle_{2}=\frac{|\Phi^+\rangle|\Phi^-\rangle+|\Phi^-\rangle|\Phi^+\rangle}{2}\\
\{(I\otimes \sigma_x),(\sigma_z\otimes i\sigma_y)\}:|\lambda\rangle_{3}= \frac{|\Psi^+\rangle|\Phi^+\rangle+|\Psi^-\rangle|\Phi^-\rangle}{2}\\
\{(I\otimes i\sigma_y),(\sigma_z\otimes \sigma_x)\}: |\lambda\rangle_{4}=\frac{|\Psi^+\rangle|\Phi^-\rangle+|\Psi^-\rangle|\Phi^+\rangle}{2}\\
\{(\sigma_x\otimes I),(i\sigma_y\otimes \sigma_z)\}: |\lambda\rangle_{5}=\frac{|\Psi^+\rangle|\Phi^+\rangle-|\Psi^-\rangle|\Phi^-\rangle}{2}\\
\{(\sigma_x\otimes \sigma_z),(i\sigma_y\otimes I)\}: |\lambda\rangle_{6}=\frac{|\Psi^+\rangle|\Phi^-\rangle-|\Psi^-\rangle|\Phi^+\rangle}{2}\\
\{(\sigma_x\otimes \sigma_x),(i\sigma_y\otimes i\sigma_y)\}: |\lambda\rangle_{7}=\frac{|\Phi^+\rangle|\Phi^+\rangle-|\Phi^-\rangle|\Phi^-\rangle}{2}\\
\{(\sigma_x\otimes i\sigma_y),(i\sigma_y\otimes \sigma_x)\}: |\lambda\rangle_{8}=\frac{|\Phi^+\rangle|\Phi^-\rangle-|\Phi^-\rangle|\Phi^+\rangle}{2}\\
\end{split}
\end{equation}
where $|\Phi^\pm\rangle=(|00\rangle\pm|11\rangle)/\sqrt[]{2}$ and $|\Psi^\pm\rangle=(|01\rangle\pm|10\rangle)/\sqrt[]{2}$ are the Bell states. Alice and Bob then send their qubits to Charlie. At this point, all four qubits are in possession of second party, i.e., Charlie and Dennis (first \& second qubits with Charlie and third \& fourth with Dennis). These states now take the form $\{|\lambda\rangle^{CCDD}_i;i=1,2,3...8\}$. Now, Charlie is supposed to decode the secrets (the sequence of operations performed by Alice and Bob).
\subsubsection*{\textbf{Step-4: Bob sends a classical bit to Charlie}}
\label{step4}
There arises a degeneracy in the states corresponding to different sequences of operations performed by Alice and Bob. We propose to resolve this issue of identifying unique operations performed by Bob by making him send a classical bit to Charlie and thus, revealing him the operation he performed. This is a\textit{ two-end }secure scheme being carried out only between Bob and Charlie. There are four values of this classical bit corresponding to $I,\sigma_x,i\sigma_y$ and $\sigma_z$. Bob encodes this classical bit into a 4-level qudit (d=4). We use a general teleportation scheme for this purpose as proposed in \cite{ref11}. Consider the 4-level qudit
\begin{equation}
|\chi_B\rangle=\sum_{k=0}^3\alpha_k|B_k\rangle
\end{equation}
in possession of Bob where the classical bit is encoded as one of the $\alpha_k$'s which can be detected by a defined projector in $\mathbb{C}^4$ as a measurement device. Teleportation is performed by the action of a global operator $(|\psi\rangle\langle\psi|_{BC}\otimes \mathbb{I}_{C'})$ where, $|\psi\rangle_{CC'}=\frac{1}{2}\sum_{i=1}^3|C_i\rangle\otimes|C'_i\rangle$ is a maximally entangled state in the system $CC'$, $B$ denotes Bob's possession and, $C$ and $C'$ denote systems in Charlie's possession. They perform the operation:
\begin{equation}
(|\psi\rangle\langle\psi|_{BC}\otimes \mathbb{I}_{C'})(|\chi\rangle_B\otimes|\psi\rangle_{CC'})
\end{equation}
\begin{equation}
=\frac{1}{8}\sum_{k,l=0}^3\alpha_k|B_l\rangle_B\otimes|A_l\rangle_C\otimes|B_k\rangle_{C'}
\end{equation}
\begin{equation}
=\frac{1}{4}|\psi\rangle_{BC}\otimes|\chi\rangle_{C'}
\end{equation}
Hence, the state, $|\chi\rangle_B$, which contained the information of Bob's local operation is now with Charlie in the same state $|\chi\rangle_{C'}$. Thus, Charlie measures this state to conclude what Bob initially had operated on his qubit. This will leave us with four possible states out of eight which were possible otherwise. This teleportation scheme is implemented to avoid the eavesdropping from any outsider who might temper with the classical bit to get Bob's measurement result and partially detect the states. If eavesdropping takes place, Charlie may detect this by contacting Bob via classical channel and discard this particular run of the protocol and start all over again. More security is guaranteed because eavesdropping cannot be successful without knowing the classical results of Dennis who is localized in party-2. Therefore, high security is maintained.
\subsubsection*{\textbf{Step-5: Dennis performs a Bell measurement on his qubits}}
\label{step5}
Finally, Dennis performs a Bell measurement in the basis $\{|\Phi^+\rangle, |\Phi^-\rangle, |\Psi^+\rangle, |\Psi^-\rangle \}_{DD}$. Charlie now needs to collaborate with Dennis to know his measurement result by getting to know from him which operation he performed. If Dennis agrees to collaborate then it will enable Charlie to decode the secret. To make this possible Dennis sends two classical bits $\{00,01,10,11\}$ to Charlie corresponding to his measurement basis. $00$ implies Dennis operated with the projector $|\Phi^+\rangle\langle\Phi^+|_{DD}$ and so on. 
Suppose, Charlie received the state $|B_4\rangle$ from Bob corresponding to $\alpha_4$. This would mean that Bob performed $\sigma_z$ on his qubit and leaves us with the possible states $|\lambda\rangle_1,|\lambda\rangle_2,|\lambda\rangle_5$ and $|\lambda\rangle_6$. Now each state is uniquely identified if Dennis performs a Bell measurement on his qubits. If Dennis sends 00 to Charlie then the state would be $|\lambda\rangle_1$, $|\lambda\rangle_2$ for 01, $|\lambda\rangle_5$ for 10 and $|\lambda\rangle_6$ for 11. Evidently, the state of Charlie's qubits will now acquire one of the Bell states! Bell states can be perfectly distinguished by performing the required projection measurement. Hence, with this scheme Charlie can uniquely determine the secret encoded by Alice as she performed local operations on her qubit at the start of the protocol.

\section{Secret sharing under energy dissipation}
\label{sec:noisy protocol}
Now we implement our scheme when Alice and Bob send their qubits to Charlie through an amplitude damping channel with varying noise. Rest of the functionality of the protocol prior to this step remains the same. Our channel has the following Kraus operators:
\begin{equation}
E_1=|0\rangle\langle0|+\sqrt[]{1-\gamma} |1\rangle\langle1|\\ 
\end{equation}
\begin{equation}
E_2=\sqrt[]{\gamma}|0\rangle\langle1|
\end{equation}
where $\gamma$ is the channel noise parameter and $0\leq\gamma\leq1$. Clearly, $E_1^\dagger E_1+E_2^\dagger E_2=\mathbb{I}$, that is, they follow \textit{completeness} relation. $\gamma$ is associated with the probability of an excited state (like that of a Harmonic Oscillator or Photon in a cavity (Jaynes-Cummings Model)\cite{ref12}) to lose energy and fall into the ground state. $\gamma$ can also be expressed as a function of the decoherence time $\tau$. We start now from step-3 of the protocol in the \textbf{noisy environment}.

\subsection*{\textit{step-3: States after Bob operates locally on his qubit}}
At this stage, the states are still pure and not subjected to the open-environmental noise. So, the density operator of the state $|\lambda\rangle_i$ is $\rho_i$, where $\rho_i=|\lambda\rangle_{ii}\langle\lambda|$ for $i=1,2,3,...8$.

Alice and Bob now send their qubits through the channels described by these Kraus operators to Charlie. This implies that the operator-sum representation will act on the first and second qubits. For a single qubit, the final state $\rho'$ after the noisy channel action, is given as,
\begin{equation}
\rho'=\sum_iE_i\rho E_i^\dagger
\end{equation}
where $E_i$ are the Kraus operators.We apply amplitude damping channel on the state $\rho_i$ (on first two qubits) as:
\begin{equation}
\begin{split}
\rho_i'=(E_1 \otimes E_1 \otimes I \otimes I)\rho_i(E_1 \otimes E_1 \otimes I \otimes I)^\dagger\\+(E_1 \otimes E_2 \otimes I \otimes I)\rho_i(E_1 \otimes E_2 \otimes I \otimes I)^\dagger\\+(E_2 \otimes E_1 \otimes I \otimes I)\rho_i(E_2 \otimes E_1 \otimes I \otimes I)^\dagger\\+(E_2 \otimes E_2 \otimes I \otimes I)\rho_i(E_2 \otimes E_2 \otimes I \otimes I)^\dagger
\end{split}
\end{equation}
which gives (8 states which are each $16\times16$ matrices in $\mathbb{C}^{16\times16}$),
\begin{equation}
\begin{split}
\rho_1'=\frac{1}{2}|0000\rangle\langle0000|+\frac{1-\gamma}{2}|1111\rangle\langle0000|
+\frac{\gamma^2}{2}|0011\rangle\langle0011|\\+\frac{(1-\gamma)\gamma}{2}|0111\rangle\langle0111|
+\frac{(1-\gamma)\gamma}{2}|1011\rangle\langle1011|\\+\frac{1-\gamma}{2}|0000\rangle\langle1111|
+\frac{1-2\gamma+\gamma^2}{2}|1111\rangle\langle1111|             
\end{split}
\end{equation}
\begin{equation}
\begin{split}
\rho_2'=\frac{1}{2}|0000\rangle\langle0000|-\frac{1-\gamma}{2}|1111\rangle\langle0000|
+\frac{\gamma^2}{2}|0011\rangle\langle0011|\\+\frac{(1-\gamma)\gamma}{2}|0111\rangle\langle0111|
+\frac{(1-\gamma)\gamma}{2}|1011\rangle\langle1011|\\-\frac{1-\gamma}{2}|0000\rangle\langle1111|
+\frac{1-2\gamma+\gamma^2}{2}|1111\rangle\langle1111|             
\end{split}
\end{equation}
\begin{equation}
\begin{split}
\rho_3'=\frac{\gamma}{2}|0000\rangle\langle0000|+\frac{\gamma}{2}|0011\rangle\langle0011|+\\\frac{1-\gamma}{2}|0100\rangle\langle0100|
+\frac{1-\gamma}{2}|1011\rangle\langle0100|\\+\frac{1-\gamma}{2}|0100\rangle\langle1011|
+\frac{1-\gamma}{2}|1011\rangle\langle1011|\\            
\end{split}
\end{equation}
\begin{equation}
\begin{split}
\rho_4'=\frac{\gamma}{2}|0000\rangle\langle0000|+\frac{\gamma}{2}|0011\rangle\langle0011|+\\\frac{1-\gamma}{2}|0100\rangle\langle0100|
-\frac{1-\gamma}{2}|1011\rangle\langle0100|\\-\frac{1-\gamma}{2}|0100\rangle\langle1011|
+\frac{1-\gamma}{2}|1011\rangle\langle1011|\\            
\end{split}
\end{equation}
\begin{equation}
\begin{split}
\rho_5'=\frac{\gamma}{2}|0000\rangle\langle0000|+\frac{\gamma}{2}|0011\rangle\langle0011|+\\\frac{1-\gamma}{2}|0111\rangle\langle0111|
+\frac{1-\gamma}{2}|1000\rangle\langle0111|\\+\frac{1-\gamma}{2}|0111\rangle\langle1000|
+\frac{1-\gamma}{2}|1000\rangle\langle1000|\\            
\end{split}
\end{equation}
\begin{equation}
\begin{split}
\rho_6'=\frac{\gamma}{2}|0000\rangle\langle0000|+\frac{\gamma}{2}|0011\rangle\langle0011|+\\\frac{1-\gamma}{2}|0111\rangle\langle0111|
-\frac{1-\gamma}{2}|1000\rangle\langle0111|\\-\frac{1-\gamma}{2}|0111\rangle\langle1000|
+\frac{1-\gamma}{2}|1000\rangle\langle1000|\\            
\end{split}
\end{equation}
\begin{equation}
\begin{split}
\rho_7'=\frac{\gamma^2}{2}|0000\rangle\langle0000|+\frac{1}{2}|0011\rangle\langle0011|
+\frac{\gamma-1}{2}|1100\rangle\langle0011|\\+\frac{(1-\gamma)\gamma}{2}|0100\rangle\langle0100|
+\frac{(1-\gamma)\gamma}{2}|1000\rangle\langle1000|\\+\frac{\gamma-1}{2}|0011\rangle\langle1100|
+\frac{1-2\gamma+\gamma^2}{2}|1100\rangle\langle1100|\\         
\end{split}
\end{equation}
\begin{equation}
\begin{split}
\rho_8'=\frac{\gamma^2}{2}|0000\rangle\langle0000|+\frac{1}{2}|0011\rangle\langle0011|
-\frac{\gamma-1}{2}|1100\rangle\langle0011|\\+\frac{(1-\gamma)\gamma}{2}|0100\rangle\langle0100|
+\frac{(1-\gamma)\gamma}{2}|1000\rangle\langle1000|\\-\frac{\gamma-1}{2}|0011\rangle\langle1100|
+\frac{1-2\gamma+\gamma^2}{2}|1100\rangle\langle1100|\\
\end{split}
\end{equation}
Amongst these states, $\rho_1', \rho_2', \rho_7', \rho_8'$ have a symmetry and states $\rho_3', \rho_4', \rho_5', \rho_6'$ are symmetric in the number of terms and coefficients. These are all density matrices with unit trace. Now, all the qubits aided with noise are in possession of Charlie and Dennis as in the case of pure state protocol. 
\subsection*{\textit{Step-4: Bob sends a classical bit to Charlie}}
This step precisely remains the same as earlier case of ideal channel. And we assume that this teleportation takes place without noise and is entirely a process involving only Bob and Charlie. Furthermore, we take a specific case when Bob measures $\sigma_z$ and sends the classical bit corresponding to $|B_4\rangle$, i.e., $\alpha_4$. Therefore, now Charlie knows that the possible states are $\rho_1', \rho_2', \rho_5'$ and $\rho_6'$ (by same analogy from the ideal case). Charlie now waits for Dennis to perform a measurement and send him two classical bits corresponding to Dennis' measurement. Rest of the three cases can be similarly explored when Bob uses $I,\sigma_x$ or $i\sigma_y$.

\subsection*{\textit{Step-5: Dennis performs a two qubit measurement}}
Now, Dennis may perform a measurement on his two qubits. There are 4 possible general elements of operation which would give the possibility of total 32 possible states forming 4 pairs based on Dennis' measurement. We define the general measurement operators as:
\begin{equation}
\begin{split}
P_1=(\alpha|00\rangle+\beta|11\rangle)(\alpha|00\rangle+\beta|11\rangle)^\dagger \\
P_2=(\alpha|00\rangle-\beta|11\rangle)(\alpha|00\rangle-\beta|11\rangle)^\dagger \\
Q_1=(\alpha|01\rangle+\beta|10\rangle)(\alpha|01\rangle+\beta|10\rangle)^\dagger \\
Q_2=(\alpha|01\rangle-\beta|10\rangle)(\alpha|01\rangle-\beta|10\rangle)^\dagger \\
\end{split}
\end{equation}
with $|\alpha|^2+|\beta|^2=1$ and $\alpha,\beta\in \mathbb{C}$. These measurements maybe considered as Bell measurements in a rotated basis. Bell measurement would now correspond to $\alpha=\beta=\frac{1}{\sqrt[]{2}}$. Dennis, thus, sends Charlie 00 for $P_1$, 01 for $P_2$, 10 for $Q_1$ and 11 for $Q_2$. Let us look at the case where he uses $P_1$ and sends 00 to Charlie (rest of the cases have similar symmetry, hence won't be mentioned).

States after the measurement by Dennis are calculated as (for example):
\begin{equation}
\rho_i^{00}=tr_{3,4}[(I\otimes I \otimes P_1)\rho_i'(I\otimes I \otimes P_1)^\dagger]
\end{equation}
for the state $\rho_i'$ (later normalized), and same follows for other states, where $tr_{i,j}(.)$ denotes the partial trace wrt sub-systems $i$ and $j$.
After the operation of $P_1$ on last two qubits under Dennis' possession and tracing out Dennis' sub-system we are left with the following states (recall from the last step, that $\sigma_z$ was applied by Bob):
\begin{equation}
\rho_1^{00}=
\begin{bmatrix}
|\alpha|^2+|\beta|^2\gamma^2 & 0 & 0 & -\alpha^*\beta(\gamma-1) \\
0 & -|\beta|^2(\gamma-1)\gamma & 0 & 0 \\
0 & 0 & -|\beta|^2(\gamma-1)\gamma & 0 \\
-\alpha\beta^*(\gamma-1) & 0 & 0 & |\beta|^2(\gamma-1)^2 \\
\end{bmatrix}
\end{equation}
\begin{equation}
\rho_2^{00}=
\begin{bmatrix}
|\alpha|^2+|\beta|^2\gamma^2 & 0 & 0 & \alpha^*\beta(\gamma-1) \\
0 & -|\beta|^2(\gamma-1)\gamma & 0 & 0 \\
0 & 0 & -|\beta|^2(\gamma-1)\gamma & 0 \\
\alpha\beta^*(\gamma-1) & 0 & 0 & |\beta|^2(\gamma-1)^2 \\
\end{bmatrix}
\end{equation}
\begin{equation}
\rho_5^{00}=
\begin{bmatrix}
\gamma & 0 & 0 & 0 \\
0 & -|\beta|^2(\gamma-1) & -\alpha\beta^*(\gamma-1) & 0 \\
0 & -\alpha^*\beta(\gamma-1) & -|\alpha|^2(\gamma-1) & 0 \\
0 & 0 & 0 & 0 \\
\end{bmatrix}
\end{equation}\\
\begin{equation}
\rho_6^{00}=
\begin{bmatrix}
\gamma & 0 & 0 & 0 \\
0 & -|\beta|^2(\gamma-1) & \alpha\beta^*(\gamma-1) & 0 \\
0 & \alpha^*\beta(\gamma-1) & -|\alpha|^2(\gamma-1) & 0 \\
0 & 0 & 0 & 0 \\
\end{bmatrix}
\end{equation}
Notice that these states are almost lying in the orthogonal sub-spaces  except for $|00\rangle\langle00|$ element which is common in all these states. Now in the final step, Charlie is supposed to distinguish between these states via some possible optimization technique.

\subsection*{\textit{Step-6: Charlie detects the right state through optimization}}
Charlie now has all the required information from Bob and Dennis using which he can decrypt the secret (the operation applied by Alice). Now he applies two projectors $\{M_1=|00\rangle\langle00|+|11\rangle\langle11|, M_2=|01\rangle\langle01|+|10\rangle\langle10|\}$ with $M_1+M_2=\mathbb{I}$, to classify these four states into two classes.Furthermore, he looks for an optimized POVM within these classes of states which gives the least error rate of distinction for them \cite{ref4}.
\\

\textbf{Classification under projector $M_1$}:\\ 
First we look at the class constructed by the application of $M_1$. Each possible state is calculated as $\rho_i^{M_1}=M_1\rho_i^{00}M_1^\dagger$. This gives us three states ($\rho_5^{00}$ and $\rho_6^{00}$ go to the same state under this projector hence, they cannot be distinguished under this class):
\begin{equation}
\begin{split}
\rho_1^{M_1}=\frac{|\alpha|^2+|\beta|^2\gamma^2}{2}|00\rangle\langle00|
+\frac{(1-\gamma)\alpha^*\beta}{2}|00\rangle\langle11|\\
+\frac{(1-\gamma)\alpha\beta^*}{2}|11\rangle\langle00|+\frac{|\beta|^2(\gamma-1)^2}{2}|11\rangle\langle11|
\end{split}
\end{equation}
\begin{equation}
\begin{split}
\rho_2^{M_1}=\frac{|\alpha|^2+|\beta|^2\gamma^2}{2}|00\rangle\langle00|
-\frac{(1-\gamma)\alpha^*\beta}{2}|00\rangle\langle11|\\
-\frac{(1-\gamma)\alpha\beta^*}{2}|11\rangle\langle00|+\frac{|\beta|^2(\gamma-1)^2}{2}|11\rangle\langle11|
\end{split}
\end{equation}
\begin{equation}
\rho_{5}^{M_1}=\frac{\gamma}{2}|00\rangle\langle00|
\end{equation}
\begin{equation}
\rho_{6}^{M_1}=\frac{\gamma}{2}|00\rangle\langle00|
\end{equation}
\\ \\
These are unnormalized states which occur with probabilities:
\begin{equation}
\begin{split}
\eta_1=\frac{|\alpha|^2+|\beta|^2((\gamma-1)^2+\gamma^2)}{2}
\end{split}
\end{equation}
\begin{equation}
\begin{split}
\eta_2=\frac{|\alpha|^2+|\beta|^2((\gamma-1)^2+\gamma^2)}{2}
\end{split}
\end{equation}
\begin{equation}
\begin{split}
\eta_{5}=\frac{\gamma}{2}
\end{split}
\end{equation}
\begin{equation}
\begin{split}
\eta_{6}=\frac{\gamma}{2}
\end{split}
\end{equation}
respectively. If $\gamma=0$, then states $\rho_{5}^{M_1}$ and $\rho_{6}^{M_1}$ would not occur in this class and this case reduces to the noiseless one. These states can be treated as \textit{state preparation} and as shown in \textbf{Theorem-3} of \cite{ref13}, the lower bound on the \textbf{inconclusive probability} $P_0^1$ is given by:
\begin{equation}
P_0^1 \geq \sqrt[]{\frac{n}{n-1}\sum_{i\neq j}\eta_i \eta_j F^2(\rho_i^{M_1},\rho_j^{M_1})}
\end{equation}
where $F(\rho,\sigma)$ is the fidelity between $\rho$ and $\sigma$ (here n=4).

It can be shown that maximum distinction is obtained when $\alpha=\beta=1/\sqrt[]{2}$, that is, when Dennis uses the Bell measurement instead of a general measurement in a rotated basis. This may be obtained by a similar technique as in \cite{ref4}. From this point, we present the results when Bell measurement is used.
\begin{figure}[hbtp]
\centering
\includegraphics[scale=0.145]{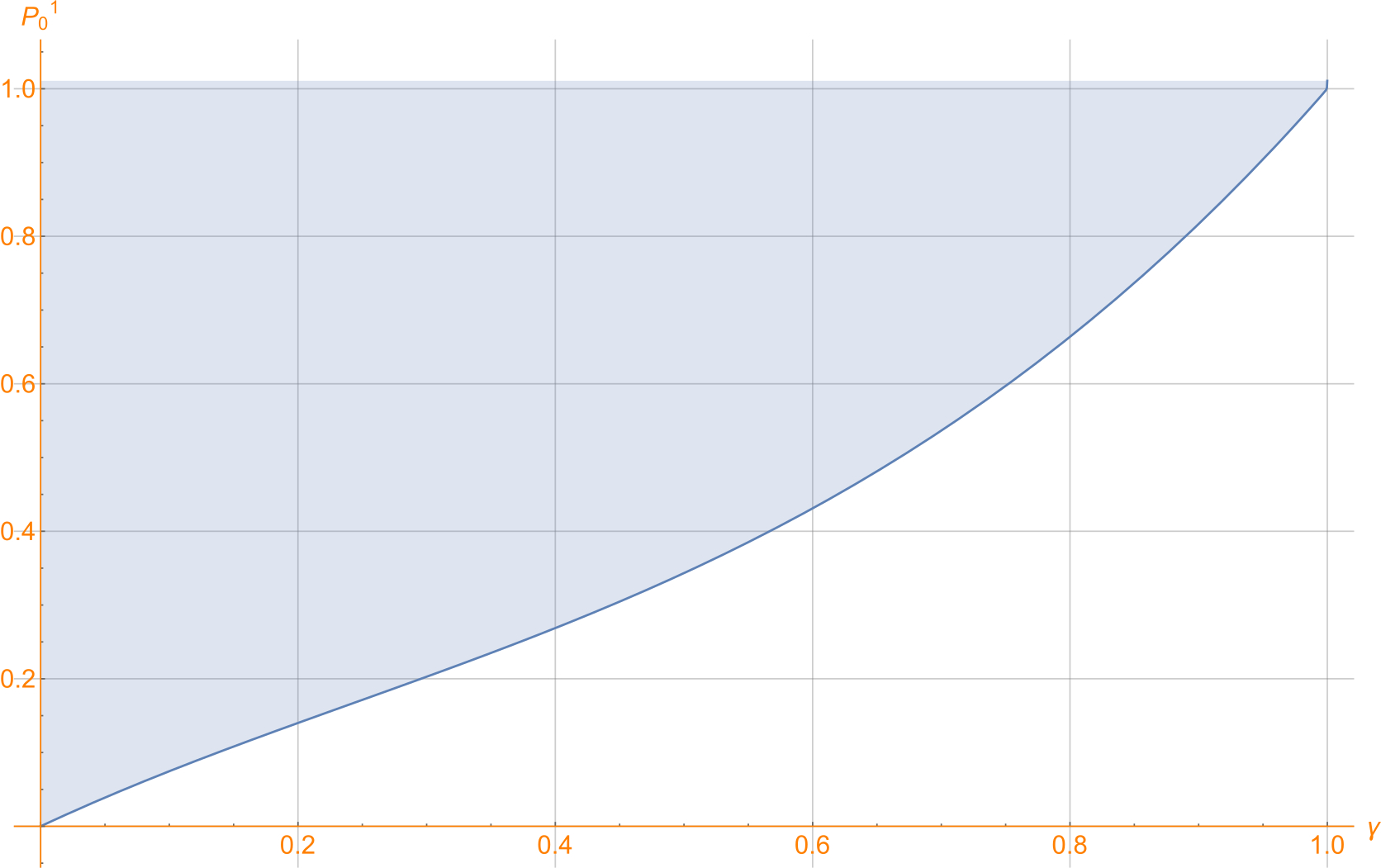}
\caption{$P_0^1$ w.r.t $\gamma$ under the class defined by $M_1$ when $\alpha=\beta=1/\sqrt[]{2}$. The shaded region is the solution for $P_0^1$ with $P_0^1\leq1$.}
\label{fig:cl1}
\end{figure}\\

\textbf{Finding the optimized POVM under the classification by $M_1$}:\\ 
We now look for an optimized POVM set $\{U_i\}$ which can distinguish the states $\{\rho_1^{M_1}, \rho_2^{M_1}, \rho_5^{M_1}, \rho_6^{M_1}\}$ as done in \cite{ref4}. Noticing the symmetry in the density matrices in this class and their relative trace weights, we propose one such POVM as:
\begin{equation}
\begin{split}
U_1=\frac{1}{2}
\begin{bmatrix}
1-2u & 1 \\
1 & 1 \\
\end{bmatrix}
,U_2=\frac{1}{2}
\begin{bmatrix}
1-2u & -1 \\
-1 & 1 \\
\end{bmatrix},
\\
U_5=
\begin{bmatrix}
u & 0 \\
0 & 0 \\
\end{bmatrix}
,U_6=
\begin{bmatrix}
u & 0 \\
0 & 0 \\
\end{bmatrix}
\\
\end{split}
\end{equation}
for some $u\in[0,1]$. This is set of valid POVM since $\sum_{i=1,2,5,6}U_i=\mathbb{I}$ and are positive semi-definite matrices for a domain of $u$ that may be chosen. Now we calculate the error rate $E_r^1$ via this POVM:
\begin{equation}
E_r^1=1-\frac{1}{2}tr[U_1\rho_1^{M_1}+U_2\rho_2^{M_1}+U_5\rho_5^{M_1}+U_6\rho_6^{M_1}]
\end{equation}
\begin{equation}
=\frac{1}{2}(1+\frac{1-\gamma}{2}-\frac{(1-\gamma)^2}{4}-u\gamma-\frac{1}{4}(1-2u)(1+\gamma^2))
\end{equation}
We plot this function in a 3D-plot.
\begin{figure}[hbtp]
\centering
\includegraphics[scale=0.13]{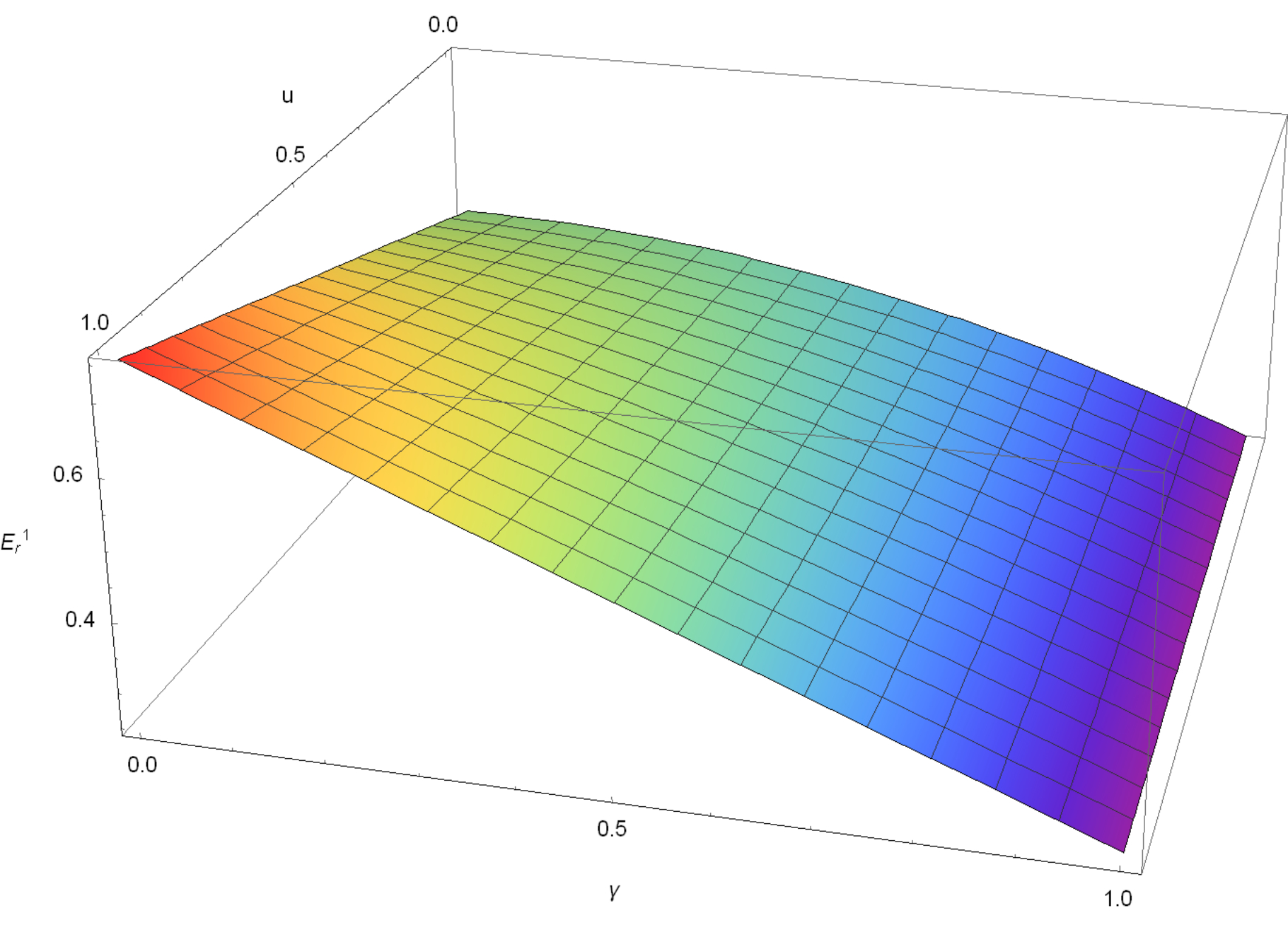}
\caption{$E_r^1$ as a function of $u$ and $\gamma$.}
\end{figure}\\

\textbf{Classification under projector $M_2$}:\\ 
Secondly, we explore the second classification, under the action of $M_2$. Each possible state is calculated as $\rho_i^{M_2}=M_2\rho_i^{00}M_2^\dagger$. This provides us with three states ($\rho_1^{00}$ and $\rho_2^{00}$ go to the same state under this projector and hence they cannot be distinguished under this class):
\begin{equation}
\begin{split}
\rho_1^{M_2}=\frac{|\beta|^2\gamma(1-\gamma)}{2}(|01\rangle\langle01|
+|10\rangle\langle10|)
\end{split}
\end{equation}
\begin{equation}
\begin{split}
\rho_2^{M_2}=\frac{|\beta|^2\gamma(1-\gamma)}{2}(|01\rangle\langle01|
+|10\rangle\langle10|)
\end{split}
\end{equation}
\begin{equation}
\begin{split}
\rho_{5}^{M_2}=\frac{|\beta|^2(1-\gamma)}{2}|01\rangle\langle01|
+\frac{\alpha\beta^*(1-\gamma)}{2}|01\rangle\langle10|\\
+\frac{\alpha^*\beta(1-\gamma)}{2}|10\rangle\langle01|+\frac{|\alpha|^2(1-\gamma)}{2}|10\rangle\langle10|
\end{split}
\end{equation}
\begin{equation}
\begin{split}
\rho_{6}^{M_2}=\frac{|\beta|^2(1-\gamma)}{2}|01\rangle\langle01|
-\frac{\alpha\beta^*(1-\gamma)}{2}|01\rangle\langle10|\\
-\frac{\alpha^*\beta(1-\gamma)}{2}|10\rangle\langle01|+\frac{|\alpha|^2(1-\gamma)}{2}|10\rangle\langle10|
\end{split}
\end{equation}
These are unnormalized states which occur with the probabilities:
\begin{equation}
\begin{split}
\zeta_1=\frac{|\beta|^2\gamma(1-\gamma)}{2}
\end{split}
\end{equation}
\begin{equation}
\begin{split}
\zeta_2=\frac{|\beta|^2\gamma(1-\gamma)}{2}
\end{split}
\end{equation}
\begin{equation}
\begin{split}
\zeta_{5}=\frac{1-\gamma}{2}
\end{split}
\end{equation}
\begin{equation}
\begin{split}
\zeta_{6}=\frac{1-\gamma}{2}
\end{split}
\end{equation}
respectively. If $\gamma=0$, then states $\rho_{1}^{M_2}$ and $\rho_{2}^{M_2}$ would not occur in this class and this case reduces to the noiseless one. Similarly, states can be treated as a state preparation and the lower bound on the \textbf{inconclusive probability} $P_0^2$ is given by:
\begin{equation}
P_0^2 \geq \sqrt[]{\frac{n}{n-1}\sum_{i\neq j}\zeta_i \zeta_j F^2(\rho_i^{M_2},\rho_j^{M_2})}
\end{equation}
(here also n=4).

\begin{figure}[hbtp]
\centering
\includegraphics[scale=0.145]{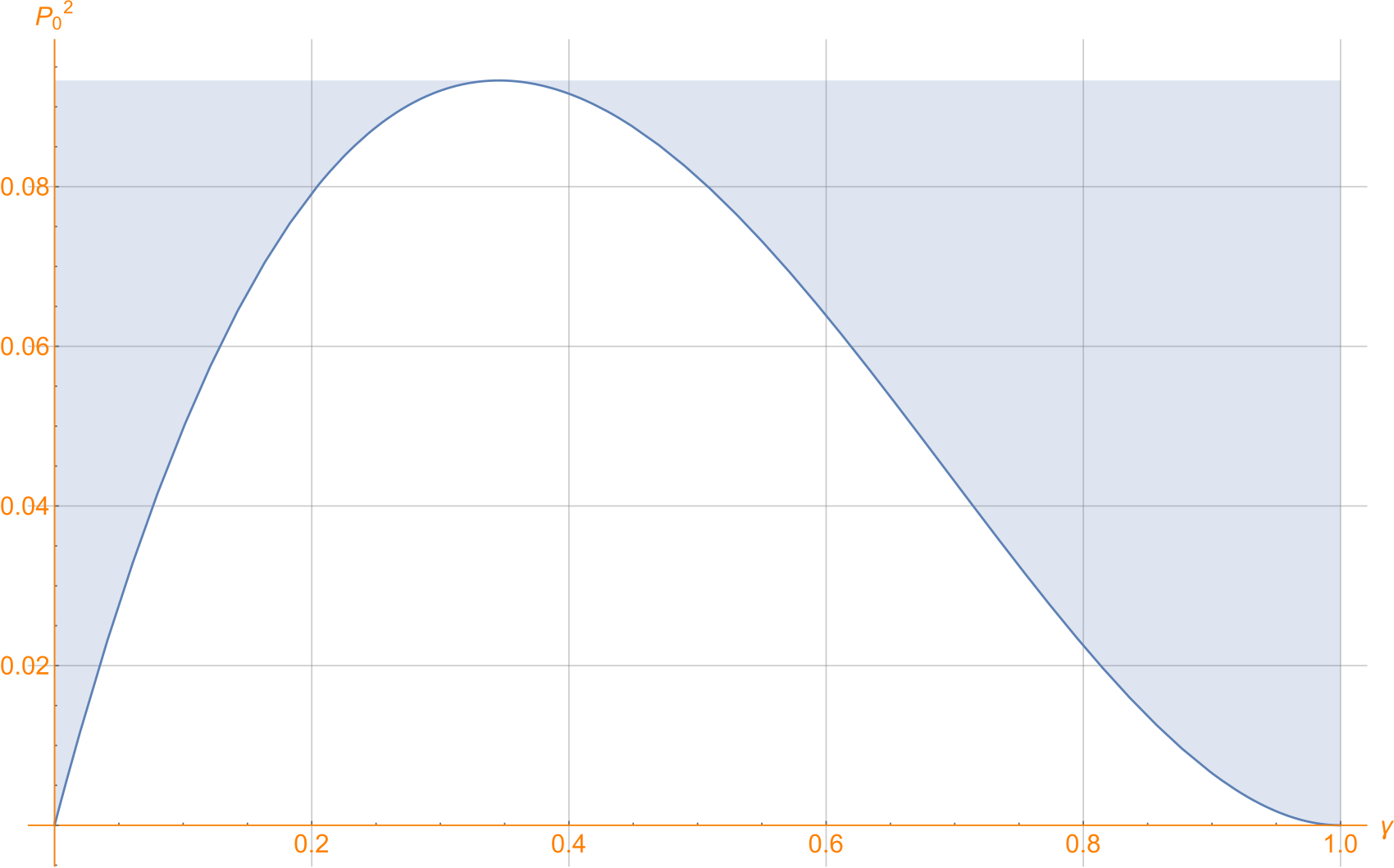}
\caption{$P_0^1$ w.r.t $\gamma$ under the class defined by $M_2$ when $\alpha=\beta=1/\sqrt[]{2}$. The shaded region is the solution for $P_0^2$ with $P_0^2\leq1$.}
\label{fig:p02}
\end{figure}
As evident from the plots for $P_0^1$ and $P_0^2$, there is an asymmetry for distinction between the two classifications w.r.t $\gamma$. This might be altered for bringing a particular scheme for distinction, by the use of a rotated basis instead of a Bell basis by Dennis and maintaining a constant level of noise in the channels.\\

\textbf{Finding the optimized POVM under the classification by $M_2$}:\\ 
We now look for an optimized POVM set $\{V_i\}$ which can distinguish the states $\{\rho_1^{M_2}, \rho_2^{M_2}, \rho_5^{M_2}, \rho_6^{M_2}\}$. Using the same symmetry in the density matrices in this class and their relative trace weights, we propose one such POVM as:
\begin{equation}
\begin{split}
V_1=
\begin{bmatrix}
0 & v \\
v & 0 \\
\end{bmatrix}
,V_2=
\begin{bmatrix}
0 & v \\
v & 0 \\
\end{bmatrix},
\\
V_5=\frac{1}{2}
\begin{bmatrix}
1 & 1-2v \\
1-2v & 1 \\
\end{bmatrix}
,V_6=\frac{1}{2}
\begin{bmatrix}
1 & -1-2v \\
-1-2v & 1 \\
\end{bmatrix}
\\
\end{split}
\end{equation}
for some $v\in[0,1]$. This is set of valid POVM since $\sum_{i=1,2,5,6}V_i=\mathbb{I}$ and are positive semi-definite matrices for a domain of $v$ that may be chosen. Now we calculate the error rate $E_r^2$ via this POVM:
\begin{equation}
E_r^2=1-\frac{1}{2}tr[V_1\rho_1^{M_2}+V_2\rho_2^{M_2}+V_5\rho_5^{M_2}+V_6\rho_6^{M_2}]
\end{equation}
\begin{equation}
=\frac{1}{4}(1-(1-\gamma)(1-2v-2v\gamma)+\gamma)
\end{equation}
We plot this function in a 3D-plot.
\begin{figure}[hbtp]
\centering
\includegraphics[scale=0.13]{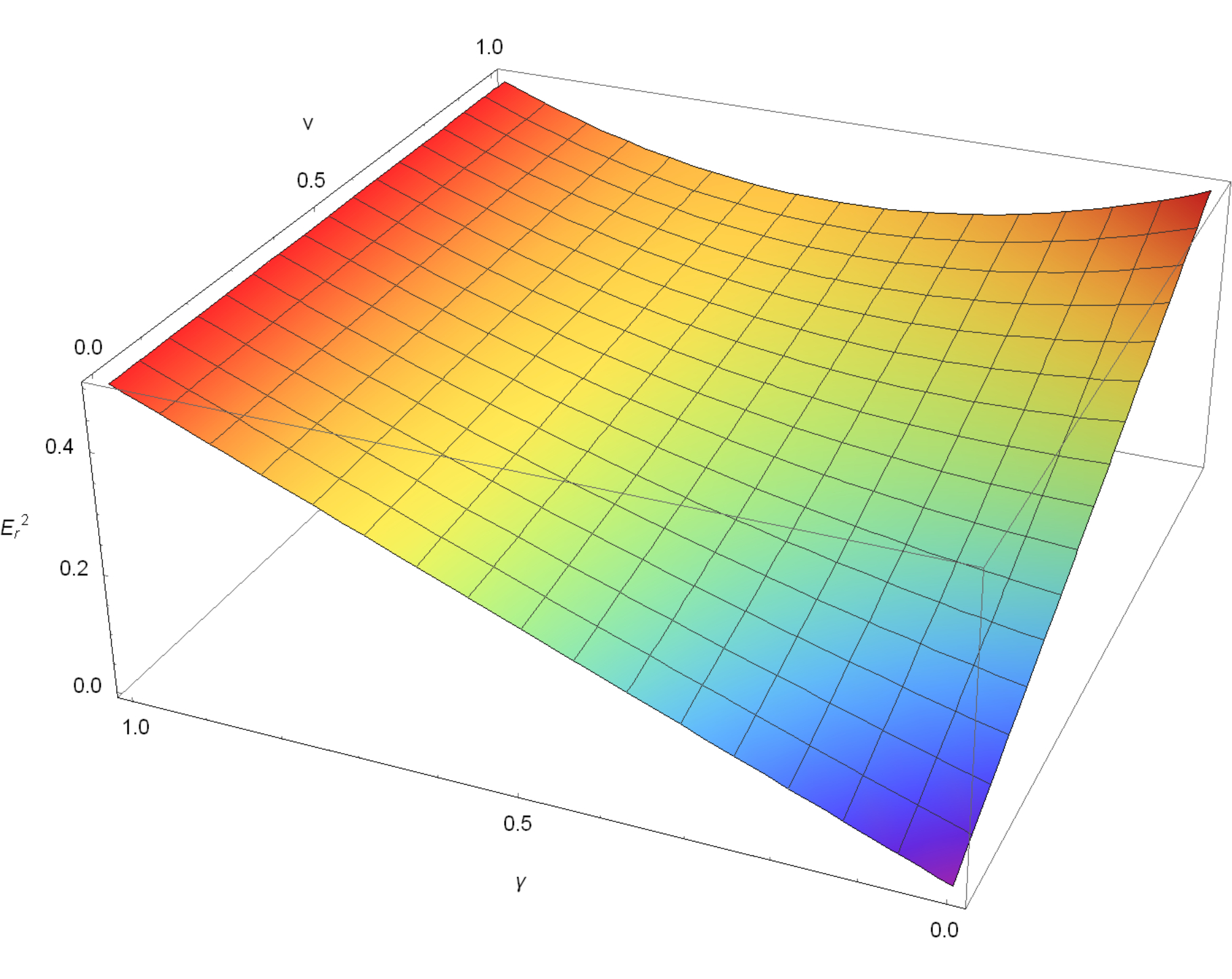}
\caption{$E_r^2$ as a function of $v$ and $\gamma$.}
\end{figure}\\

These graphs for $E_r^1$ and $E_r^2$ give the geometric solution for $u$ and $v$ for a given value of $\gamma$ such that the bounds of $P_0^1$ and $P_0^2$ are satisfied (plotted for a Bell measurement). We now calculate the total number of bits $T_{bits}$ that Charlie can decode (1 from Bob, 2 from Dennis and 1 from Alice by all over error minimization).
\begin{equation}
T_{bits}=4-(E_r^1+E_r^2)
\end{equation}
A triplet $(\gamma,u,v)$ can be found from these graphs satisfying the bound for $P_0^1$ and $P_0^2$. Or at least, $u$ and $v$ may be found for a given value of $\gamma$ that lets Charlie decode maximum number of bits. A better optimization technique may guarantee a much stronger upper bound for the total number of bits that Charlie can decode.
\section{Channel quality measures}
\label{sec:quality}
In this section, we look at various measures that convey the quality of the channel as a function of its decoherence factor $\gamma$. We look at \textit{Fidelity} between the initial state and the resultant state after the action of the channel and then two measures of \textit{Coherence}, specifically, the $l_1$-\textit{norm of coherence} and \textit{relative entropy of coherence}.\\

\textbf{Fidelity of the states}:\\
We calculate the fidelity of all the eight states after the measurements were done by Alice and Bob. Fidelity is calculated between their pure forms and the respective mixed states after the action of the channel as:
\begin{equation}
F_i=F(|\lambda \rangle _i,\rho_i')=\sqrt[]{_i\langle\lambda|\rho_i'|\lambda\rangle_i}
\end{equation}
This gives the fidelities as:
\begin{equation}
F_1=F_2=F_7=F_8=\sqrt[]{1-\gamma+\frac{\gamma^2}{4}}
\end{equation}
and,
\begin{equation}
F_3=F_4=F_5=F_6=\sqrt[]{1-\gamma}
\end{equation}
\begin{figure}[hbtp]
\centering
\includegraphics[scale=0.25]{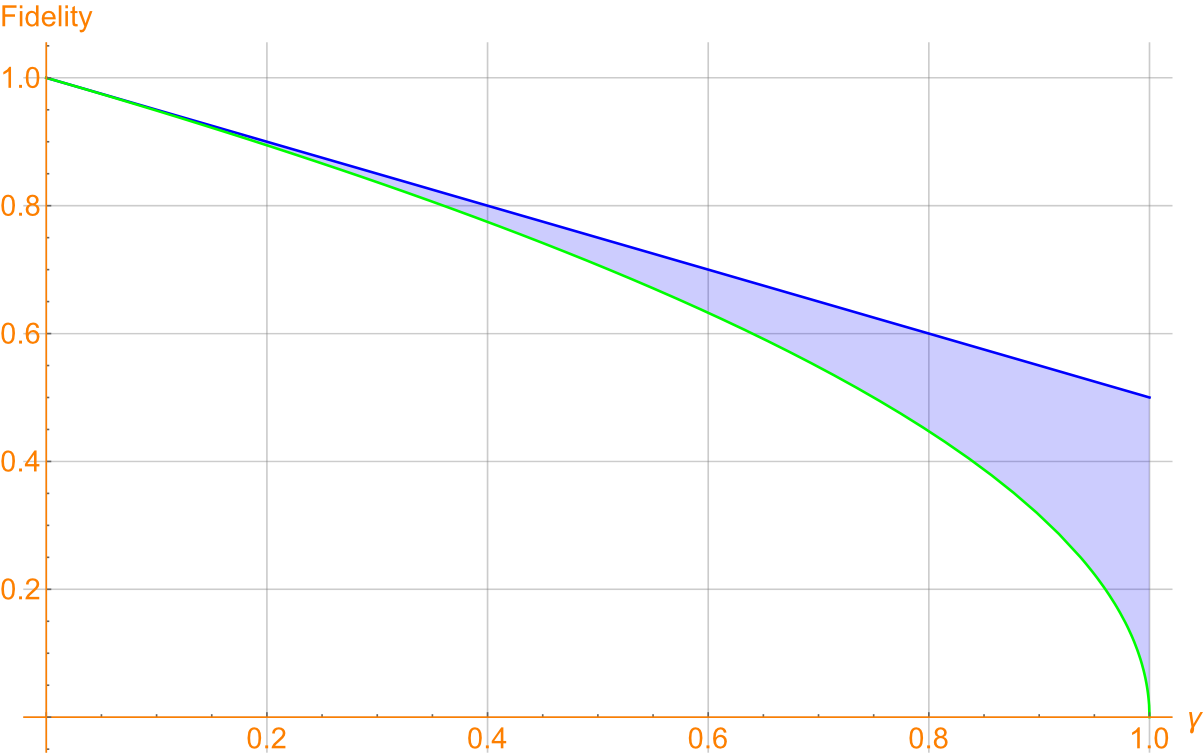}
\caption{Fidelity measure w.r.t $\gamma$. \textbf{Blue line:} Fidelity of states 1,2,7 \& 8. \textbf{Green line:}Fidelity of states 3,4,5 \& 6.}
\label{fig:fidelity}
\end{figure}
Evident from the plot that as the noise increases, the fidelity is lost. Thus, the channel is associated with more loss of information and hence less ability to distinguish in such a scenario.\\

\textbf{$l_1$ norm of Coherence}:
Given a density matrix $\rho$, the $l_1$ norm of coherence is given as:
\begin{equation}
C_{l_1}(\rho)=\sum_{i\neq j}|\rho_{i,j}|
\end{equation}
This gives,
\begin{equation}
C_{l_1}(\rho_i')=|1-\gamma|, \forall i\in \{1,2,3...8\}
\end{equation} 
\begin{figure}[hbtp]
\centering
\includegraphics[scale=0.16]{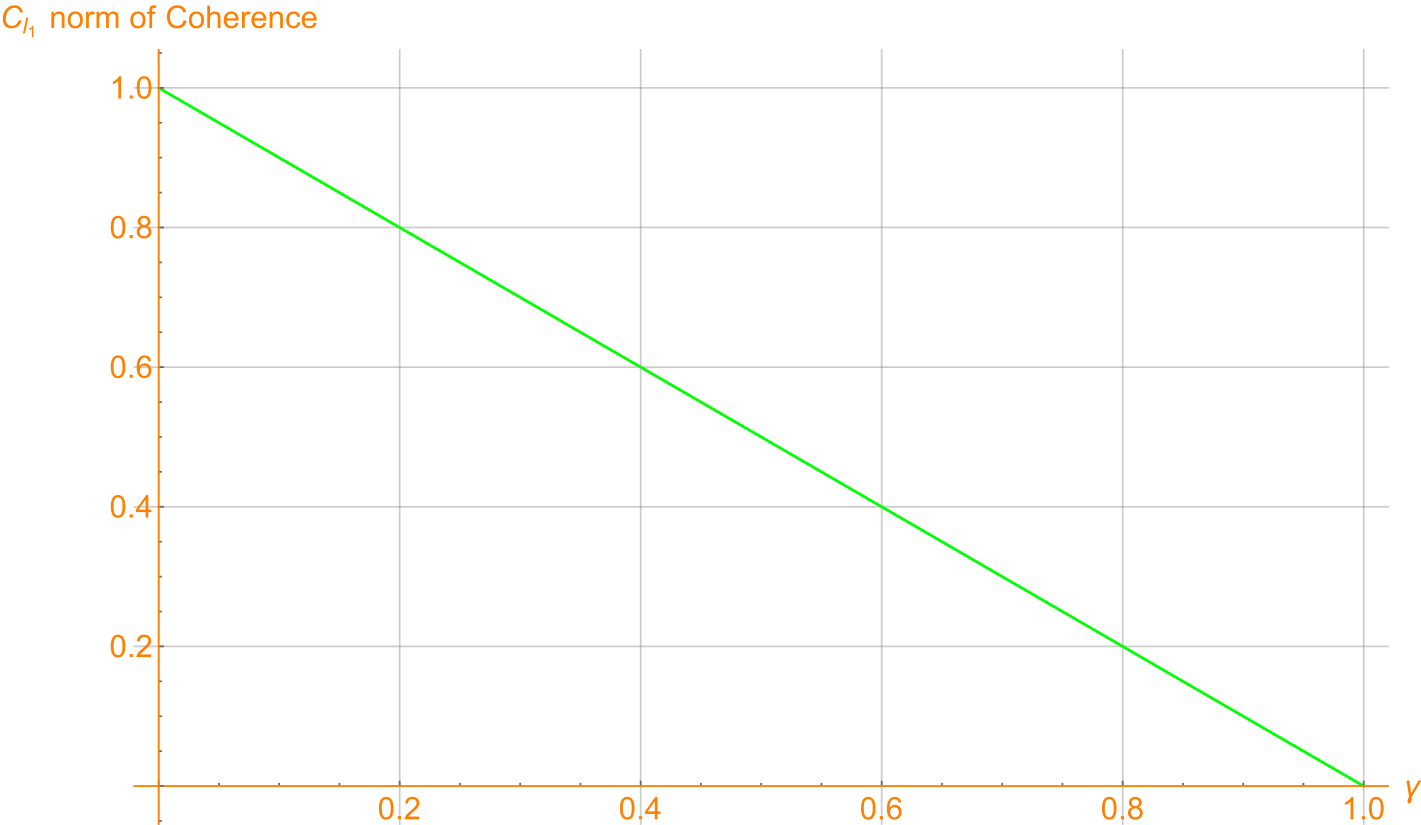}
\caption{$C_{l_1}$ norm of coherence w.r.t $\gamma$ for all the 8 possible states after the action of the channel.}
\label{fig:cl1}
\end{figure}\\

\textbf{Relative entropy of Coherence}:\\
Relative entropy of Coherence is calculated as:
\begin{equation}
C_r(\rho_i')=S(\rho_{i,diag}')-S(\rho_i')
\end{equation}
where $S(\rho)=-\sum_j\lambda_j\log_2(\lambda_j)$ (here, $\lambda_j$'s are eigenvalues of $\rho$), is the von Neumann Entropy of $\rho$ and, $\rho_{diag}$ is the \textit{diagonal} form of $\rho$, when all off-diagonal elements are forced to zero that carry relative phases-\textit{coherence} (it is a decohered form of the state $\rho$). The calculation gives:
\begin{equation}
\begin{split}
C_r(\rho_1')=C_r(\rho_2')=C_r(\rho_7')=C_r(\rho_8')\\
=\frac{1}{2}-\frac{(1-\gamma)^2}{2}\log_2(\frac{(1-\gamma)^2}{2})\\
+\frac{2-2\gamma+\gamma^2}{2}\log_2(\frac{2-2\gamma+\gamma^2}{2})
\end{split}
\end{equation}
and,
\begin{equation}
\begin{split}
C_r(\rho_3')=C_r(\rho_4')=C_r(\rho_5')=C_r(\rho_6')=1-\gamma
\end{split}
\end{equation}
$C_r(\rho_i')$ is plotted w.r.t $\gamma$.\\
\begin{figure}[hbtp]
\centering
\includegraphics[scale=0.12]{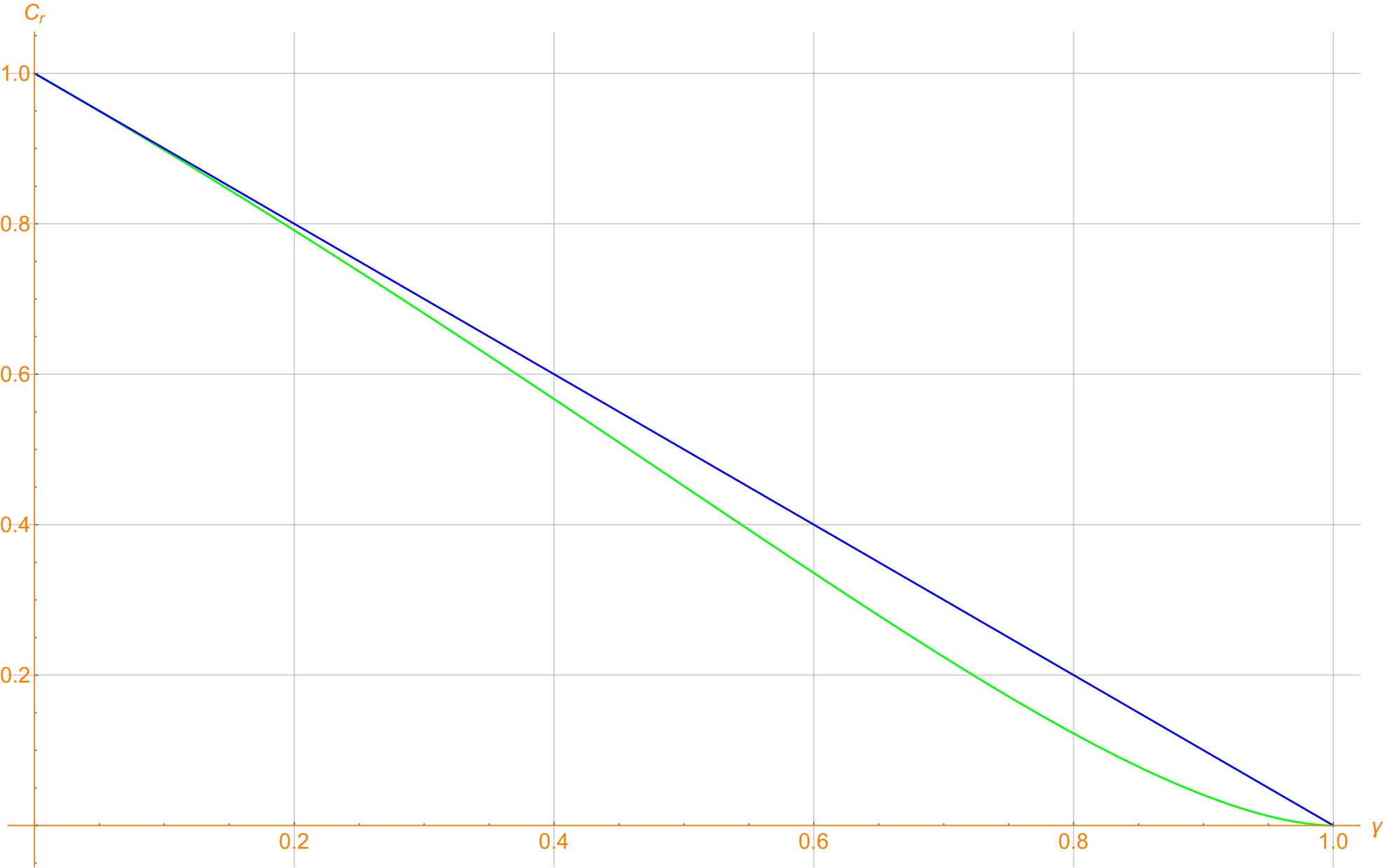}
\caption{$C_r$ norm of coherence w.r.t $\gamma$. \textbf{Green line:} $C_r$ for states $\rho_1'$, $\rho_2'$, $\rho_7'$ \& $\rho_8'$. \textbf{ Blue line:} $C_r$ for states $\rho_3'$, $\rho_4'$, $\rho_5'$ \& $\rho_6'$.}
\label{fig:cr}
\end{figure}\\
It is clear from the plots that coherence is lost as $\gamma$ is increased, which means that, the states get rapidly decohered. Hence, the value of $\gamma$ must be maintained in the domain where least \textit{quantumness} of states is lost with respect to a minimum noise trade off! Various entanglement witness and measures also have same nature under such a channel. Under local operations and classical communication (LOCC), it cannot increase but may decrease on the repeated local measurements. This can be seen from the fact that the GHZ state taken no longer retains its maximal entanglement.
\section*{Conclusion}
\label{sec:conclusion}
We presented a protocol which uses four maximally entangled GHZ qubits as a correlation resource. We used this to implement a multipartite secret sharing which can be extended to a many body network using an analogous scheme. This is a scheme much closer to the real world situation where noise cannot be perfectly eliminated. Hence, the state distinction becomes very difficult because the channel depletes the fidelity and coherence of the state. While tackling the problem of finding an \textit{optimized POVM} for the mixed states to be distinguished, we went through the methods proposed in \cite{ref14} and \cite{ref15} which are applicable to our scheme, however, are hard to compute. We obtained four density matrices with non-orthogonal subspaces to be distinguished and there existed no projector which could project them onto orthogonal sub-spaces. These states were not trivial to be distinguished, such as by methods similar to the one proposed in \cite{ref4}, which was performed for a \textit{phase damping} channel and does not affect the diagonal elements of the density matrix. So, to overcome this problem, we initially followed the procedure of purification of mixed-state density matrices by \textit{purifying} (see H.J.W Theorem \cite{ref17}) them to higher dimensional Hilbert space ($\mathbb{C}^{16}$ in this case) and constructing four linearly independent vectors in $\mathbb{C}^{16}$ to be distinguished. This is a computationally hard problem if the procedure of \cite{ref14} is followed for complex vector spaces of higher dimensions (this may be achieved by exterior algebra in higher dimensions). We still are looking for better optimization techniques and finding an upper bound to the maximum number of total bits that Charlie can decode. This protocol can be implemented for \textit{depolarizing} and \textit{dephasing} channels too. Optimization for these channels will be analogous to the one in \cite{ref4}. However, we have not made the teleportation between Bob and Charlie subject to noise. This can be modeled as in \cite{ref16}. For high security of the protocol, the asymmetric distribution of the GHZ state maybe made probabilistic (sometimes sending qubits to Charlie and sometimes to Dennis) by interchanging their role. This will switch the receiver of the classical bit sent by Bob via quantum teleportation.

\end{document}